%% file: HllqqPaper.tex
\documentclass[3p,sort&compress]{elsarticle}
\usepackage{hyperref}
\usepackage{atlasphysics}
\usepackage{subfigure}
\usepackage{lineno}
\usepackage{multirow}
\usepackage{amssymb}

\newcommand\htollnunu{\ensuremath{H\to ZZ\to\ell^+\ell^-\nu\bar{\nu}}}
\newcommand\htollqq{\ensuremath{H\to ZZ\to\ell^+\ell^- q \bar{q}}}

\newcommand\jets{\mathrm{jets}}

\newcommand\mll{\ensuremath{m_{\ell\ell}}}
\newcommand\mjj{\ensuremath{m_{jj}}}
\newcommand\mlljj{\ensuremath{m_{\ell\ell jj}}}

\def\progname#1{{\sc #1}}

\def\Figref#1{Figure~\ref{#1}} %
\def\Figsref#1{Figures~\ref{#1}} %
\def\figref#1{Fig.~\ref{#1}}   %
\def\secref#1{Section~\ref{#1}}
\newcommand\lumi{\ensuremath{1.04~\ifb}}

\def\papertitle{Search for a heavy Standard Model Higgs boson 
  in the channel $\htollqq$ using the ATLAS detector}

\journal{Physics Letters B}

\begin{document}
\begin{frontmatter}
\title{\vspace{-0.5cm} \flushright{\normalsize{CERN-PH-EP-2011-136}} \\ \center{\papertitle}}
\author{The ATLAS Collaboration}
\address{}

\begin{abstract}

A search for a heavy Standard Model Higgs~boson decaying via $\htollqq$, where $\ell=e,\mu$, is
presented.  The search is performed using a data set of $pp$ collisions at $\sqrt{s}=7\tev$,
corresponding to an integrated luminosity of \lumi\ collected in 2011 by the ATLAS detector at the
CERN LHC collider.  No significant excess of events above the estimated background is found. Upper
limits at 95\%~confidence level on the production cross section (relative to that expected from the
Standard Model) of a Higgs~boson with a mass in the range between 200 and $600\gev$ are derived.
Within this mass range, there is at present insufficient sensitivity
to exclude a Standard Model Higgs boson.
For a Higgs~boson with a mass of $360\gev$, where the sensitivity is maximal, the observed and
expected cross section upper limits are factors of 1.7 and 2.7, respectively, larger than the
Standard Model prediction.

\end{abstract}


\end{frontmatter}

\section{Introduction}

The search for the Standard Model (SM)
Higgs~boson~\cite{Englert:1964et,Higgs:1964pj,Guralnik:1964eu} is one
of the most crucial goals of the LHC physics program. Direct
searches at the CERN LEP $e^+e^-$ collider have set a lower limit of
$114.4\gev$ on the Higgs~boson mass~($\mH$) at 95\% confidence
level~(CL)~\cite{lepew}.  Searches by the CDF and D0 experiments at
the Fermilab Tevatron $p\bar{p}$ collider have explored the
Higgs~boson mass range up to $200\gev$ and exclude the region
$156\gev<\mH<177\gev$~\cite{tevhiggs}.

The higher centre-of-mass energy
($\sqrt{s}$) of the LHC enables the search to be extended to much
larger Higgs~boson masses.  Results from the 2010 run of the LHC, with
$\sqrt{s}=7\tev$ and an integrated luminosity of about $40~\ipb$, have
excluded a SM-like Higgs~boson with a cross section above
$\sim5$--20 times the SM prediction in the mass range
200--$600\gev$~\cite{atlas2010,cms2010}.
Although this mass range
is indirectly excluded at 95\%~CL by global fits to 
SM observables~\cite{lepindirect}, it is crucial to complement
such indirect limits by direct searches; further,
possible extensions to the SM can conspire to allow a heavy
Higgs~boson to be compatible with existing measurements~\cite{Peskin:2001rw}.

If $\mH$ is larger than twice the $Z$~boson mass,
$\mZ$, the Higgs~boson is expected to decay to two on-shell $Z$~bosons with a high
branching fraction~\cite{LHCHiggsCrossSectionWorkingGroup:2011ti,Djouadi:1997yw,Bredenstein:2006rh,Bredenstein:2006ha}. 
In this paper, we consider the Higgs~boson mass range
$200$--$600\gev$ and search for a SM Higgs~boson
decaying to a pair of $Z$~bosons, where one $Z$~boson
decays leptonically and the other hadronically: $\htollqq$
with $\ell\equiv e,\mu$.   
This analysis uses $\lumi$ of data recorded by the ATLAS experiment in the first half of 2011.
The statistical sensitivity of the analysis is enhanced by treating
events in which the hadronically-decaying $Z$~boson decays to $b$ quarks
as a separate subsample.
The largest background to this signal is
$Z+\jets$ production, with smaller contributions from $\ttbar$
and diboson ($ZZ$, $WZ$) production.

\section{ATLAS detector}

The ATLAS detector~\cite{atlas} consists of several subsystems.  An
inner tracking detector is immersed in a~2 Tesla magnetic field produced by a
superconducting solenoid. Charged particle position measurements are made by
silicon detectors in the pseudorapidity range $|\eta|<2.5$ and by a straw
tube tracker in the range $|\eta|<2.0$%
\footnote{ATLAS uses a right-handed coordinate system with its origin
at the nominal interaction point (IP) in the centre of the detector and
the $z$-axis coinciding with the axis of the beam pipe.
The $x$-axis points from the IP to the centre of the LHC ring,
and the $y$-axis points upward. Cylindrical coordinates ($r$,$\phi$)
are used in the transverse plane,  $\phi$ being the azimuthal angle
around the beam pipe. The pseudorapidity is defined in terms
of the polar angle $\theta$ as $\eta = - \ln \tan(\theta/2)$.
For the purpose of the electron fiducial selection, this is calculated
relative to the geometric centre of the detector; otherwise, it
is relative to the primary vertex.}
The calorimeters cover
$|\eta|<4.9$ with a variety of detector technologies. The
liquid-argon electromagnetic
calorimeter is divided into barrel ($|\eta|<1.475$) and endcap
($1.375<|\eta|<3.2$) regions.
The hadronic calorimeters (using liquid argon or scintillating tiles as
active materials) surround the electromagnetic calorimeter and cover
$|\eta|< 4.9$.
The muon
spectrometer measures the deflection of muon tracks in the field of three
large superconducting toroid magnets. It is instrumented with separate
trigger ($|\eta|<2.4$) and high-precision tracking ($|\eta|<2.7$) chambers.

\section{Data and Monte Carlo samples}

The data used in this search were recorded by the ATLAS experiment during the
2011 LHC run with $pp$ collisions at $\sqrt{s}=7\tev$.  They correspond to an
integrated luminosity of approximately \lumi\ after data quality selections to
require that all systems used in this analysis were operational.  The
data were collected using primarily single-lepton triggers with a transverse
momentum ($\pt$) threshold of $20\gev$ for electrons and $18\gev$ for muons.
The
resulting trigger criteria are about 95\% efficient in the muon channel and close to
100\% efficient in the electron channel, relative to the selection criteria described
below. Collision events are selected by requiring a reconstructed primary
vertex with at least three associated tracks with $\pt>0.4\gev$.
The average number of collisions per bunch crossing in this data
sample is about six.

The \htollqq\ signal is modelled using the \progname{powheg} Monte
Carlo (MC) event generator~\cite{powheg4,powheg5},
which calculates separately the gluon and vector-boson fusion production
mechanisms of the Higgs~boson with matrix elements up to next-to-leading
order.
Events generated with \progname{powheg} are hadronized with 
\progname{pythia}~\cite{pythia},
which in turn is interfaced via
\progname{photos}~\cite{Golonka:2005pn} to model final-state radiation and
via \progname{tauola}~\cite{Was:2000st} to simulate $\tau$ decays. 
The $\htollnunu / \ell^+\ell^-\ell^+\ell^-$
processes are also simulated
and included as part of the signal, as are $Z\ra\tau\tau$ decays.
These additional signal channels comprise $< 3\%$ of the acceptance
of this analysis.
The signal is also simulated with \progname{pythia} 
in order to estimate the systematic uncertainty due to the modelling
of the signal kinematic distributions. The total
inclusive cross sections for Higgs~boson production
with their corresponding uncertainties are taken from
Refs.~\cite{LHCHiggsCrossSectionWorkingGroup:2011ti,Harlander:2002wh,Anastasiou:2002yz,Ravindran:2003um,Anastasiou:2008tj,deFlorian:2009hc,Baglio:2010ae,Bolzoni:2010xr,Catani:2003zt,Aglietti2004,Actis2008,Ciccolini:2007jr,Ciccolini:2007ec,Djouadi:1997yw,Bredenstein:2006rh,Bredenstein:2006ha,Botje:2011sn,Alekhin:2011sk,Lai:2010vv,Martin:2009iq,Ball:2011mu}.  
The combined production cross section and decay branching ratio
for the $\htollqq$ channel ranges from $140\pm20$~fb
for $\mH=200\gev$ to $10\pm2$~fb for $\mH=600\gev$. 

Various background processes are modelled with several event
generators. The \progname{alpgen} generator~\cite{alpgen}, interfaced to
\progname{herwig}~\cite{herwig} for parton showers and hadronization, is used
to simulate $W/Z+\jets$ events. The \progname{mc@nlo} generator~\cite{mcatnlo},
interfaced to \progname{jimmy}~\cite{jimmy} for the simulation of underlying
events, is used for top quark and diboson production.
The \progname{pythia} event generator is used to produce
alternative samples of $Z+\textrm{jet}$ events to study systematic uncertainties.

The SM $ZZ$ process is an irreducible background for $H\ra ZZ$. The $qq\to ZZ$ process is modelled
using the \progname{mc@nlo} generator, which only includes contributions from on-shell $Z$~bosons.
Thus, an alternative sample produced with \progname{pythia}, calculated at leading order but
including off-shell $Z$~bosons, is used to study systematic uncertainties.  The $q\bar{q}\to ZZ$
production cross section has been calculated up to next-to-leading order in
QCD~\cite{Campbell:1999ah}.  Due to the large gluon flux at the LHC, next-to-next-to-leading order
gluon pair quark-box diagrams ($gg\to ZZ$) are significant and the cross section is scaled up by 6\%
to account for this additional contribution~\cite{Binoth:2008pr}.

\section{Reconstruction and identification of physics objects}

Electron candidates are reconstructed from energy clusters in the electromagnetic calorimeter that
have matching tracks in the inner detector.  The candidates are required to pass identification criteria based on the electromagnetic
shower shape, track quality, and
track-cluster matching~\cite{Aad:2010yt}.
Muon candidates are reconstructed by matching tracks
found in the inner detector with either full or partial tracks in the muon
spectrometer~\cite{Aad:2010yt}.  To reject cosmic rays, muon candidates must be consistent with
originating from the primary vertex.  Both electrons and muons must be isolated, defined as follows.
The transverse momenta of tracks within a cone of radius
$\Delta R \equiv \sqrt{(\Delta\eta)^2+(\Delta\phi)^2} = 0.2$ around the lepton
candidate track, excluding the candidate track itself, are summed.
This sum must be less than 10\% of the transverse
momentum of the candidate.  This cut rejects jets that would otherwise mimic an electron, as well as
leptons originating from heavy-flavour decays.  Both electrons and muons must satisfy $\pt>20\gev$
and $|\eta|<2.5$ (2.47 for electrons), and electrons must not be close to any
identified muon ($\Delta R>0.2$).

Jets are reconstructed from energy clusters in the calorimeter using an anti-$k_t$
algorithm~\cite{AntiKt} with a radius parameter $R=0.4$. The jet energies are
calibrated using $\pt$-
and $\eta$-dependent correction factors based on Monte Carlo simulation and
validated on data~\cite{jes1,jes2}.
Only jets with $\pt>25\gev$ and $|\eta|<2.5$ are
considered. A jet is rejected if an identified electron candidate is found
within $\Delta R<0.4$ to avoid double counting. 
It is also discarded if less than 75\% of the transverse momentum of
its associated tracks originates from the primary vertex; this rejects
jets that originate from other collisions in the same bunch crossing.

Missing transverse momentum, \met, caused by the presence of neutrinos in an event, is an important
characteristic to help separate signal from background, and is calculated by summing the vector
transverse momenta of all calorimeter energy clusters with $|\eta|<4.5$ and all identified
muons.

Jets which originate from $b$-quarks can be discriminated from other jets
based on the  relatively long lifetime ($c\tau\approx 450~\mathrm{\mu m}$)
of hadrons containing $b$-quarks.
This is accomplished by considering the set of tracks associated
with the jet and either reconstructing a secondary vertex from among
them, or finding tracks that have a significant
impact parameter with respect to the primary event vertex~\cite{btagnote2011}.
Information from both methods is combined into a
single discriminating variable, and a cut applied that gives an 
efficiency of about $70\%$ for identifying real $b$-jets (``$b$-tagging''),
with a light-quark jet rejection of about~50.

Corrections are applied to MC events to account for various small
differences between data and simulation observed and determined in a variety of samples, including
$(J/\psi, \Upsilon, Z)\to\ell\ell$ and $W\to\ell\nu$.  Quantities
corrected include the average number of minimum-bias events per
crossing, trigger and lepton identification efficiencies, and the
lepton energy scale and resolution.

\section{Event selection}

The first step in the event selection is to reconstruct a $Z\ra\ell\ell$
decay. Events must contain exactly two same-flavour selected leptons.
The two muons of a pair must have opposite
charge; this is not required for electrons because larger energy losses from
bremsstrahlung lead to higher charge misidentification probabilities.
The pair's invariant mass must lie within the range $76\gev<\mll<106\gev$
($\approx\mZ\pm15\gev$).

In addition to the $Z\to\ell\ell$ decay, the \htollqq\ final state contains a pair of jets resulting
from $Z\to qq$ decay and no high-$\pt$ neutrinos.  Thus, events must contain at least two jets and
satisfy $\met<50\gev$.  The latter requirement reduces mostly background from $\ttbar$ production.

About 21\% of signal events contain $b$-jets from $Z\ra\bbbar$ decay, while a $b$-jet pair is rare
($\sim 2$\%) in the dominant $Z$+jets background.  Accordingly, the analysis is divided into a ``tagged'' subchannel,
containing events with two $b$-tags, and an ``untagged'' subchannel, containing events with less
than two $b$-tags.  Events with more than two $b$-tags (approximately $3\%$ of
the data sample with $\ge 2$ jets) are rejected.

Events are then required to have at least one candidate $Z\ra qq$ decay with dijet invariant mass
satisfying $70\gev<\mjj<105\gev$ in order to be consistent with a $Z$~boson decay.  This cut is
asymmetric around the $Z$ boson mass since there are non-Gaussian tails extending to lower masses.
For untagged events, all pairs of jets formed from the three leading $\pt$ jets are considered.  All
such pairs
are retained with unit weight, leading to the possibility of multiple candidates per event (the fraction of untagged events with more than one pair
retained per event is 10--16\% for the low-$\mH$ selection and 2--5\%
for the high-$\mH$ selection).
If the
event is tagged, then the two tagged jets are used to form the dijet invariant mass and their
energies are scaled up by 5\% to take into account the average jet energy scale difference between
heavy- and light-quark jets.  The dijet invariant mass distributions before the $\mjj$ requirement
are shown in \figref{fig:mjj}.

\def\figmjj{%
   \centerline{%
\includegraphics[width=0.45\textwidth]{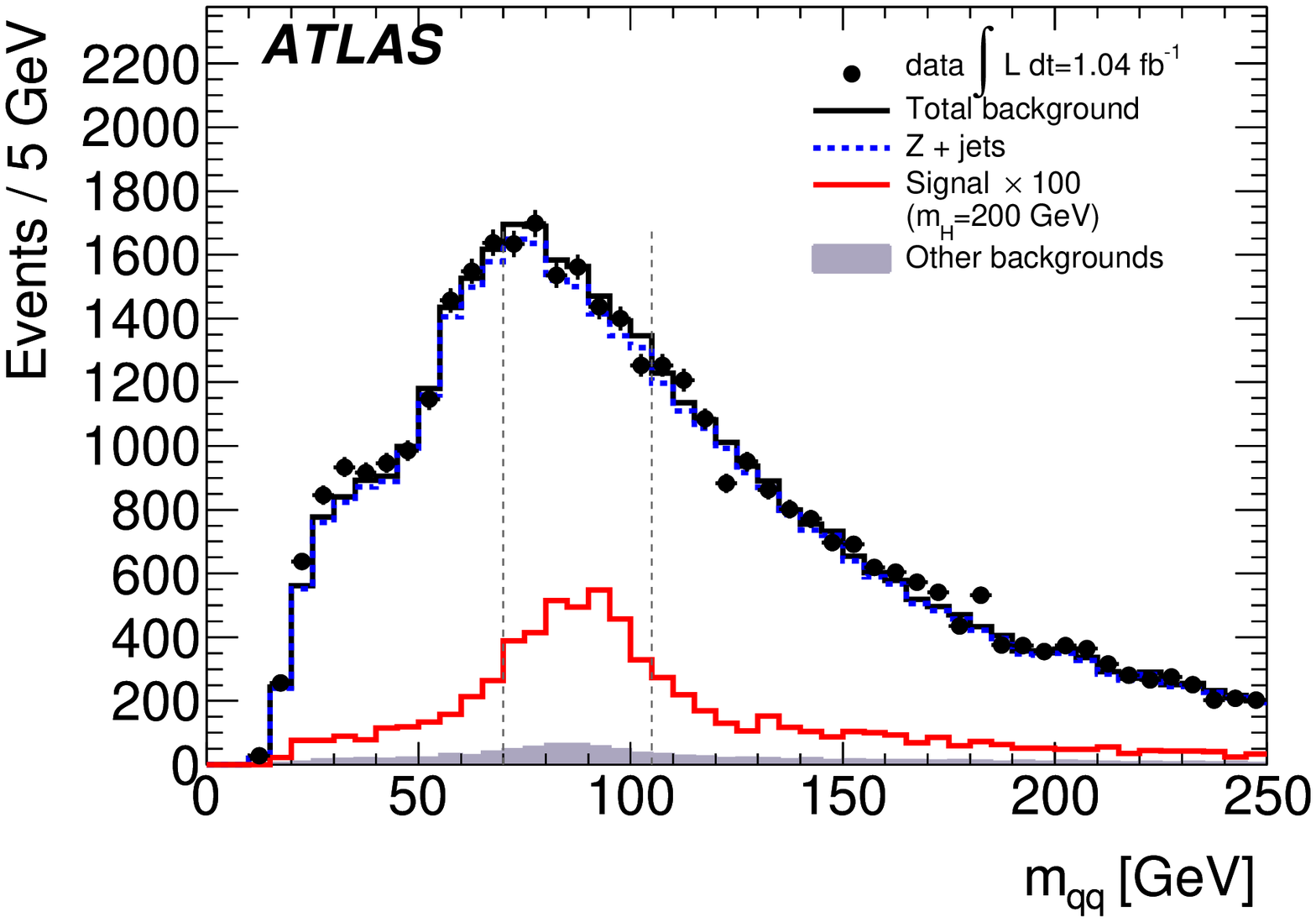}
\includegraphics[width=0.45\textwidth]{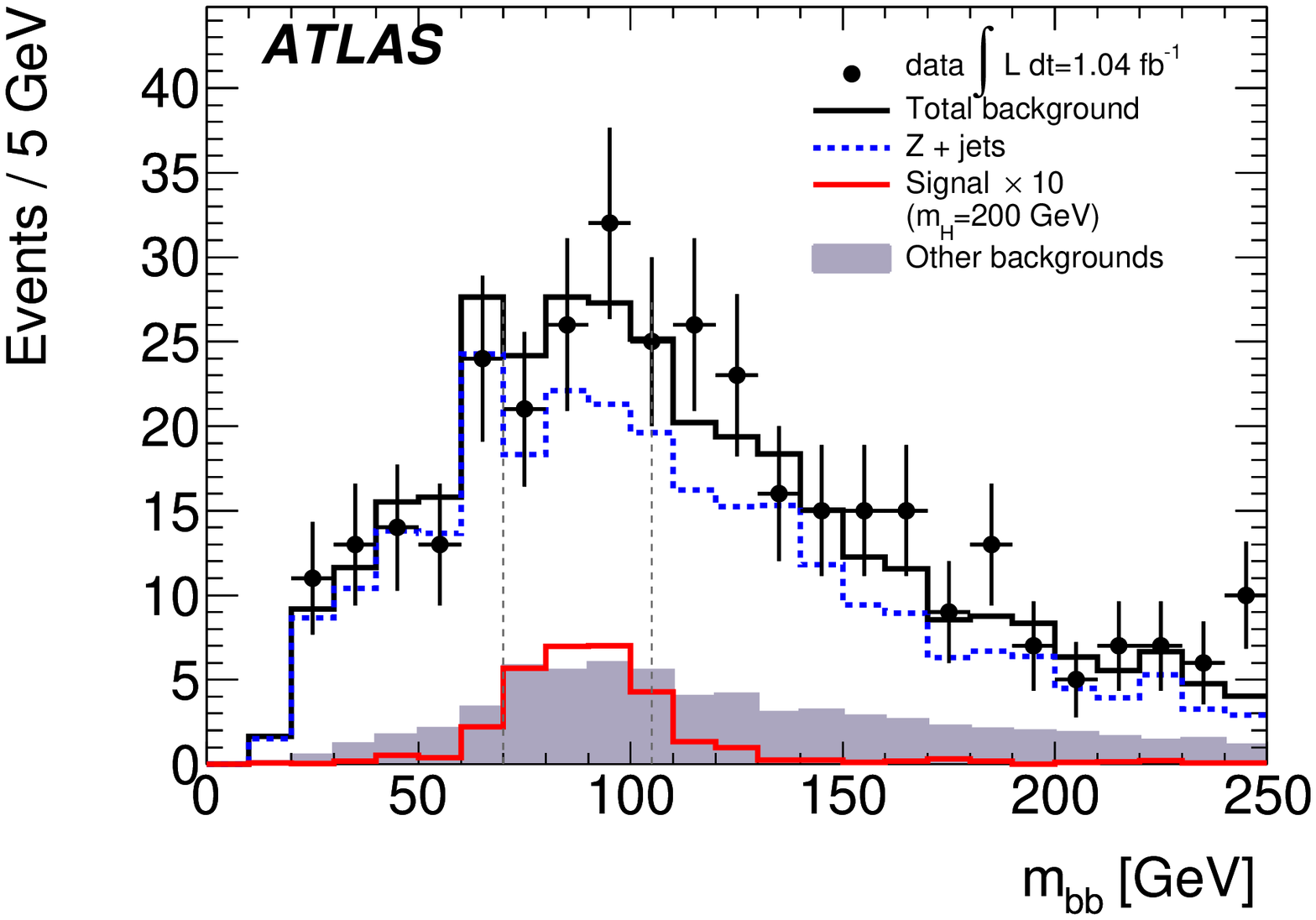}
    }
    \caption{Distributions of the invariant mass of selected dijet pairs, $\mjj$,
      for the data and the MC simulation,
      for the untagged (left) and tagged (right) samples. 
      The signal
      has been scaled up to make it more visible. 
      The vertical lines show the range of the $\mjj$
      selection.}%
}
\begin{figure}
  \figmjj
    \label{fig:mjj}
\end{figure}

These event selections define the ``low-$\mH$'' selections.  For
larger Higgs~boson masses, the $Z$~bosons from $H\to ZZ$ decays have
large momenta in the laboratory reference frame, resulting in smaller opening angles between
their decay products.  Therefore, ``high-$\mH$'' selections
are defined by the following additional requirements: (1) the two jets must
have $\pt>45\gev$, and (2) $\Delta\phi_{\ell\ell}<\pi/2$ and
$\Delta\phi_{jj}<\pi/2$.  These selections are applied 
when searching for a Higgs~boson with $\mH\ge300\gev$,
for which they improve the sensitivity.

Following this event selection, an $\htollqq$ signal is expected to appear as a peak
in the invariant mass distribution of the
$\ell\ell jj$ system, with $\mlljj$ around $\mH$.
To improve the Higgs~boson mass resolution, the energies of the jets
forming each dijet pair are scaled by a single multiplicative factor
to set the dijet invariant mass $\mjj$
to the nominal mass of the $Z$~boson.
The total efficiency for the selection of signal events
is about $13\%$ for $\mH=200\gev$ and $18\%$ for $\mH=600\gev$.

\section{Background estimates}
\label{sec:background}

The principal background to this analysis is $Z$~boson production in association with jets
($Z+\jets$).  The shape of this background is derived from \progname{alpgen} Monte Carlo simulations
and checked against data, while the normalization is derived directly from data.
\Figsref{fig:controlplots}(a) and (b) show the $\mlljj$ distribution after the jet and $\met$
requirements for events with the dijet invariant mass in sidebands of the $Z$~boson mass:
$40\gev<\mjj<70\gev$ or $105\gev<\mjj<150\gev$.  The Monte Carlo gives a good description of the shape, but
predicts about 10\% more events than are seen in the data.  The numbers of events in the sidebands,
after subtraction of the small contribution from other background sources, are used to derive scale
factors to correct the normalization of the $Z+\jets$ Monte Carlo to that observed in the data.  For
the untagged channel, scale factors are derived separately for the low- and high-$\mH$ selections;
for the tagged channel, the low-$\mH$ selection is used to derive a single scale factor, as the
tagged high-$\mH$ selection has very few events in the sidebands.  Furthermore, as the shapes
derived from the tagged \progname{alpgen} MC samples suffer from significant statistical fluctuations,
the shapes derived for the untagged selection are used for the tagged backgrounds, with appropriate
scale factors applied.  The shapes are found to agree within statistical uncertainties between the
tagged and untagged MC samples.

\def\figcontrolplots{%
   \centerline{%
\subfigure[$\mjj$ sidebands, untagged sample.]{\includegraphics[width=0.45\textwidth]{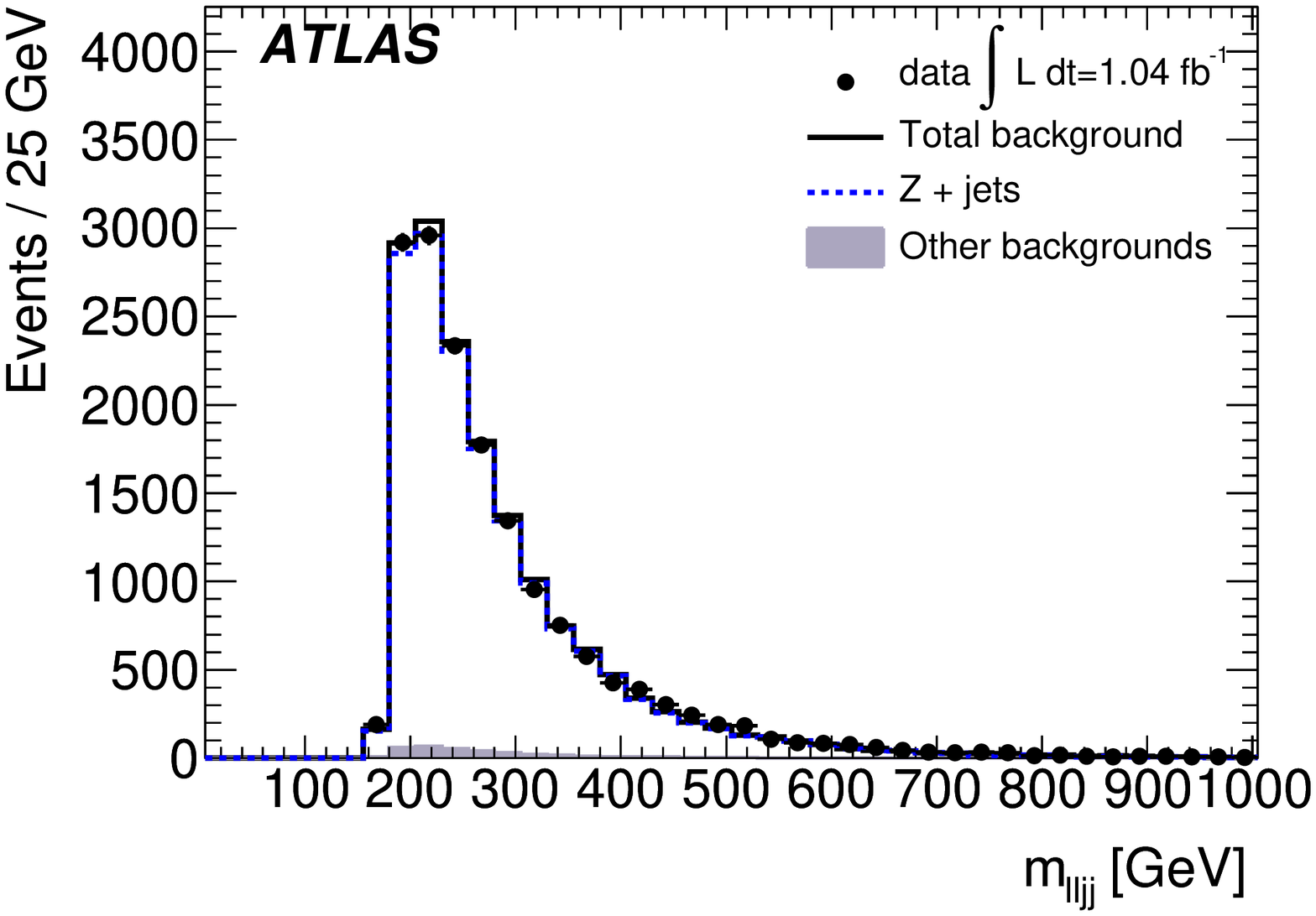}}
\subfigure[$\mjj$ sidebands, tagged sample.]{\includegraphics[width=0.45\textwidth]{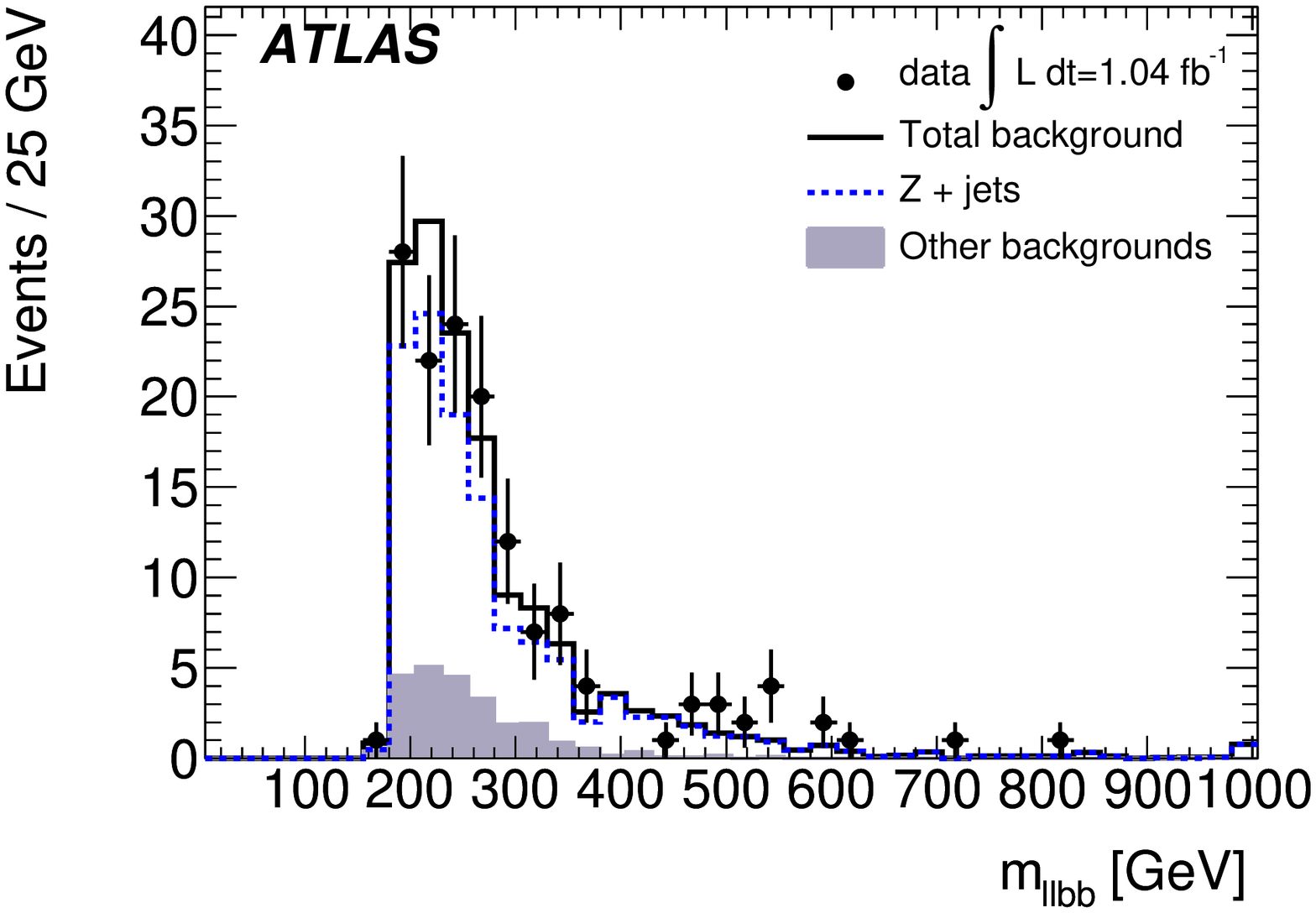}}
    }
   \centerline{%
\subfigure[$\mll$ sidebands, $\met>50\gev$, untagged sample.] {\includegraphics[width=0.45\textwidth]{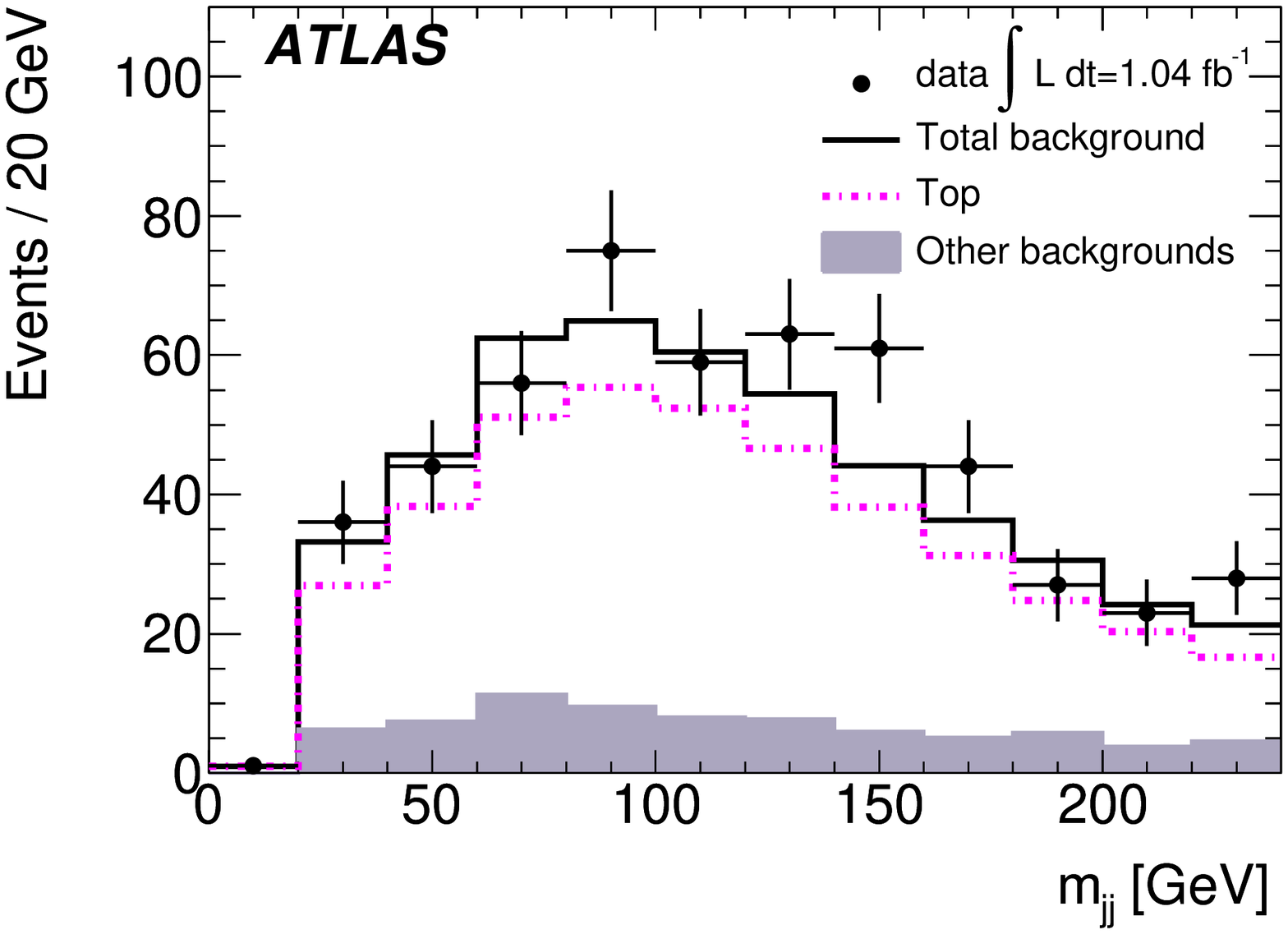}}
\subfigure[$\mll$ sidebands, $\met<50\gev$, tagged sample.] {\includegraphics[width=0.45\textwidth]{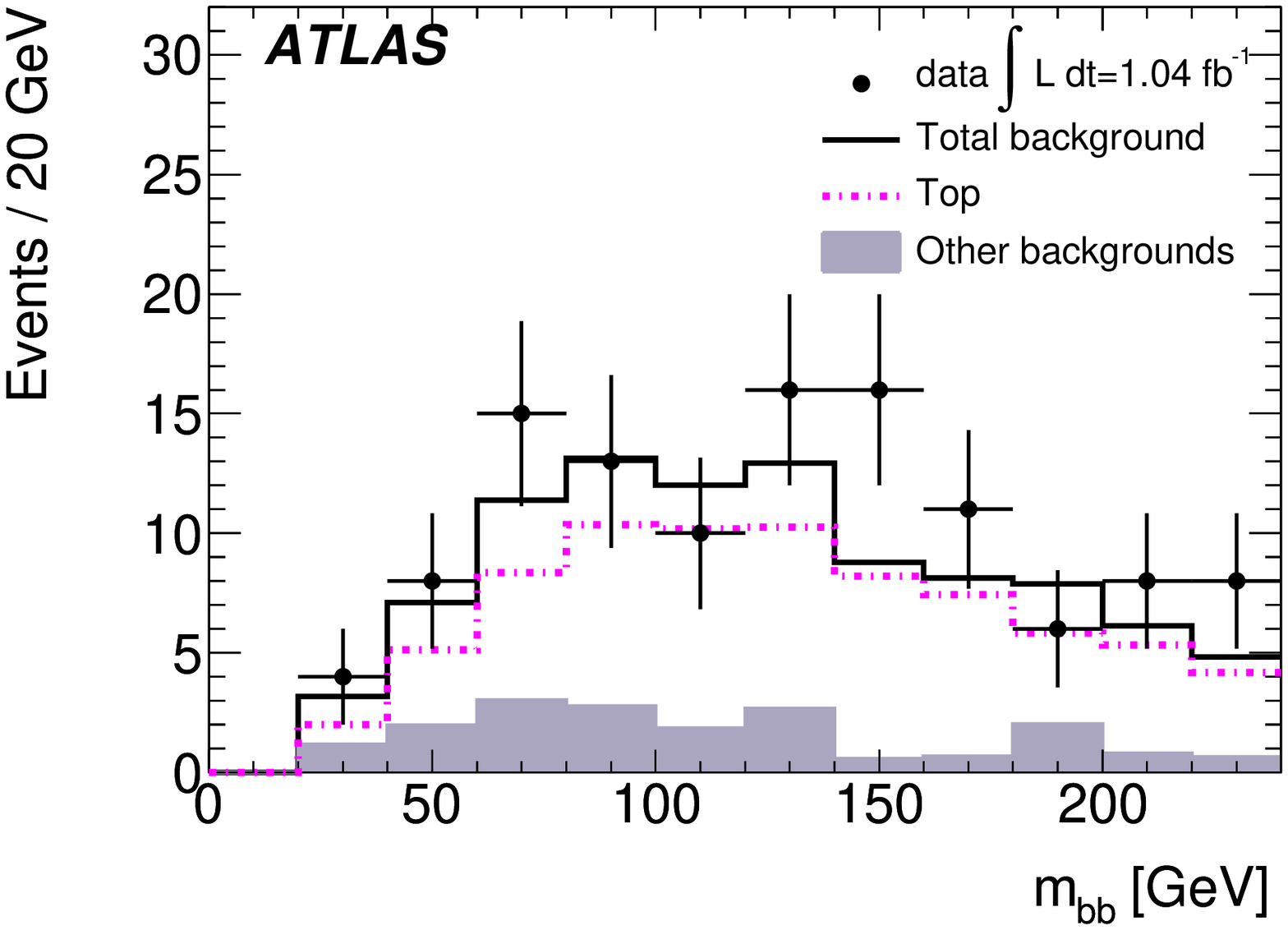}}
    }
  \caption{Distributions from the background control samples, after application
    of scale factors.  Top row: the $\ell\ell jj$ invariant mass
    for $40\gev<\mjj<70\gev$ or $105\gev<\mjj<150\gev$
    after the jet and $\met$ requirements, for (a) the untagged
    and (b) the tagged sample.
    Bottom row: the invariant mass of the $jj$ system
    for events with $60\gev<\mll<76\gev$ or
    $106\gev<\mll<150\gev$ for
    (c) the untagged sample with the additional requirement
    $\met>50\gev$ and (d) the tagged sample with $\met<50\gev$.}%
}
\begin{figure}
  \figcontrolplots
    \label{fig:controlplots}
\end{figure}

Another significant background to this analysis is top~quark production.
As for $Z+\jets$, the shape is taken from Monte Carlo and
the normalization is checked against data, using the sideband
$60\gev<\mll<76\gev$ or $106\gev<\mll<150\gev$ of the dilepton mass
distribution.  \Figsref{fig:controlplots}(c) and (d) show the $\mjj$ distributions for
these sidebands, both for the untagged selection (with the $\met$ selection
reversed) and the tagged selection.  The normalization of the $\ttbar$
component of top~quark production is calculated at NNLO using
\progname{hathor}~\cite{hathor}; for the single-top component, the
\progname{mc@nlo} normalization is used.  As the Monte Carlo
agrees with the data within uncertainties, no scale factor is applied to the simulation in this case.

The small irreducible background from $ZZ$ production is difficult to constrain from data due to the
large $Z+\jets$ background component and possible contamination from the signal.  Thus, this background
is estimated entirely from Monte Carlo simulation.  
The small backgrounds from $WZ$ and $W+\jets$ production are also taken from Monte Carlo simulation.

The background from multijet events in which jets are misidentified as isolated leptons is estimated
from data.  For the electron channel,
a sample of events is selected that contains
electron candidates that fail the selection requirements but pass loosened requirements;
the normalization is determined by a multicomponent fit to the $\mll$ distribution in events
containing at least two jets.  The multijet background in the muon channel is estimated by dividing
the dimuon + jets events into four categories based on whether the muons are isolated or
non-isolated and on whether or not the invariant mass of the muon pair lies near the $Z$~boson mass
peak.  The number of background events with two isolated muons with invariant mass consistent with
$Z$~boson decay can then be determined from the numbers of events observed in the other three
categories (which contain negligible contamination from the signal) under the assumption that the
two variables (isolation criteria and invariant mass) are uncorrelated.  The muon channel multijet background is
 found to be negligible.

\section{Systematic uncertainties}

The theoretical uncertainties on the Higgs~boson production cross
section compiled in
Ref.~\cite{LHCHiggsCrossSectionWorkingGroup:2011ti} are 15--20\% for
the gluon fusion process and 3--9\% for the vector-boson fusion
process, depending on the Higgs~boson mass\footnote{The limits
  presented in this search assume cross sections based on on-shell
  Higgs boson production and decay and use Monte Carlo generators with
  an ad-hoc Breit-Wigner Higgs~boson line shape. Potentially
  important effects related to off-shell Higgs boson production and
  interference between the Higgs boson signal and backgrounds
  have recently been discussed~\cite{LHCHiggsCrossSectionWorkingGroup:2011ti,Anastasiou:2011pi}. The
  inclusion of such effects may affect limits at very high Higgs~boson masses
  ($\mH>400\gev$).}.  Signal samples generated with \progname{pythia}
instead of \progname{powheg} are also used to evaluate the uncertainty
on the selection efficiency due to the modelling of the signal
kinematics. This results in a 3\% (6\%) uncertainty for the low-
(high-) $\mH$ selection.

The uncertainty in the normalization of the $Z+\jets$ background from
the procedure described in \secref{sec:background} is evaluated by
comparing the scale factors obtained from the upper or lower sideband
separately.  It is taken as the difference between the scale factors
or the statistical uncertainty, whichever is larger. It
is found to be 1.4\% for the low-$\mH$ untagged selection,
8.1\% for the high-$\mH$ untagged selection, and 18\% for the tagged
selections.  The uncertainty on the shapes of the $Z+\jets$ (and $ZZ$)
backgrounds is estimated using an alternate Monte Carlo sample
generated with \progname{pythia} instead of \progname{alpgen} (or
\progname{mc@nlo}).  The uncertainty on the $\ttbar$ cross section is
found by adding the contributions from variations of the QCD renormalization and factorization scales and
from the \progname{cteq6.6}~\cite{Lai:2010vv} parton distribution
function (PDF) error set; the result is 9\%.  The diboson backgrounds, which
are estimated directly from Monte Carlo, have a combined 5\% scale and
\progname{cteq6.6} PDF uncertainty on the cross section; adding an
additional 10\% uncertainty, corresponding to the maximum difference
seen between \progname{mc@nlo} and $k$-factor scaled \progname{pythia}
results, yields an overall uncertainty of 11\%.  A 100\% systematic uncertainty is assigned to the
normalization of the multijet background in the electron channel from
the procedure described in \secref{sec:background} by
comparing the result of fitting the $\mll$ distribution before and
after the requirement of at least two jets. The normalization
uncertainty for the small $W+\jets$ background is taken to be 50\%.

An overall 3.7\% uncertainty
from the total integrated luminosity~\cite{lumi2011} is added
to the uncertainties on all Monte Carlo processes (excluding $Z+\jets$,
which is normalized to data), correlated across all samples.

There are also systematic uncertainty contributions from detector effects, including the lepton and
jet trigger and identification efficiencies, the energy or momentum calibration and resolution of
the leptons and jets, and the $b$-tagging efficiency and mistag rates.  The dominant uncertainty on
the tagged sample comes from the $b$-tagging efficiency, which corresponds to an average of 16\%
(23\%) for the signal for the low- (high-) $\mH$ selection.  For the untagged sample, the uncertainty
on the jet energy scale is a major contribution, giving rise to an average uncertainty of 5\% on the
signal.

\section{Results}

Table~\ref{tab:limits} shows the numbers of candidates observed in data for each
of the four selections compared with the background expectations.
\Figref{fig:results} shows the $\mlljj$ distributions
for both the tagged and untagged channels for the low- and high-$\mH$
selections.

\begin{table}[htb]
\caption{The expected numbers of signal and background candidates
  in the $\htollqq$ channel, along with the numbers of candidates observed
  in data, for an integrated luminosity of \lumi.
  The first error indicates the statistical uncertainty,
  the second error the systematic uncertainty.}
\begingroup
\catcode`?=\active
\catcode`,=\active
\begin{center}
  {\small
    \def\cpm{$\,\pm\,$}
    \def\lcols#1{\multicolumn{3}{l}{#1}}
    \def\ccols#1{\multicolumn{3}{c}{#1}}
\newdimen\digitwidth
\newdimen\dotwidth
\setbox0=\hbox{\rm0}
\digitwidth=\wd0
\setbox0=\hbox{\rm.}
\dotwidth=\wd0
\def?{\kern\digitwidth}
\def,{\kern\dotwidth}
    \begin{tabular}{l*{4}{r@{\cpm}r@{\cpm}r}}
\hline
      & \multicolumn{6}{c}{Untagged}              & \multicolumn{6}{c}{Tagged}        \\
      & \ccols{Low-$\mH$}    & \ccols{High-$\mH$} & \ccols{Low-$\mH$}&\ccols{High-$\mH$}\\
\cline{2-13}%

$Z$+jets & 10352 & 60 & 160 &   420,? & 12,? & 30,? & 72,? & 1,? & 15,? & 4.9 & 0.2 & 1.0 \\
$W$+jets &    10 &  2 &   5 &     0.2 &  0.2 &  0.1 &  \ccols{$<0.1$}   & \ccols{$<0.1$}  \\
Top      &    40 &  1 &   6 &     3.0 &  0.3 &  0.6 & 13,? & 1,? &  3,? & 1.1 & 0.2 & 0.3 \\ 
Multijet &    64 &  3 &  60 &     2.0 &  0.5 &  2.0 &  0.3 & 0.2 &  0.3 & \ccols{$<0.1$}  \\
$ZZ$     &   107 &  4 &  15 &     8.5 &  1.1 &  1.8 &  6.9 & 1.0 &  2.0 & 0.8 & 0.2 & 0.3 \\ 
$WZ$     &   143 &  3 &  30 &    17,? &  1,? &  3,? &  0.5 & 0.2 &  0.3 & \ccols{$<0.1$} \\
\hline
Total background
         & 10718 & 60 & 170 &   450,? & 13,? & 30,? & 92,? & 1,? & 15,? & 6.9 & 0.4 & 1.2 \\
Data     &  \lcols{10495}           &  \lcols{419}             &  \lcols{91}          & \lcols{6}          \\
\hline
\multicolumn{1}{c}{Signal}\\
\cline{1--1}%
$\mH=200\gev$
         &    33 &  1 &   6 &\lcols{}                  &  2.2 & 0.2 &  0.6 &\lcols{}            \\
$\mH=300\gev$
         &\lcols{}                   &     7.0 &  0.3 &  1.5 &\lcols{}              & 0.6 & 0.1 & 0.2 \\
$\mH=400\gev$
         &\lcols{}                   &     9.8 &  0.3 &  1.8 &\lcols{}              & 1.1 & 0.1 & 0.3 \\
$\mH=500\gev$
         &\lcols{}                   &     5.5 &  0.1 &  1.0 &\lcols{}              & 0.6 & 0.0 & 0.2 \\
$\mH=600\gev$
         &\lcols{}                   &     2.5 &  0.1 &  0.5 &\lcols{}              & 0.3 & 0.0 & 0.1 \\
\hline
    \end{tabular}
  }%
\end{center}%
\endgroup
\label{tab:limits}
\end{table}

\def\figresults{%
  \begin{center}
    \subfigure[Low-$\mH$, untagged selection.]
{\includegraphics[width=0.45\textwidth]      {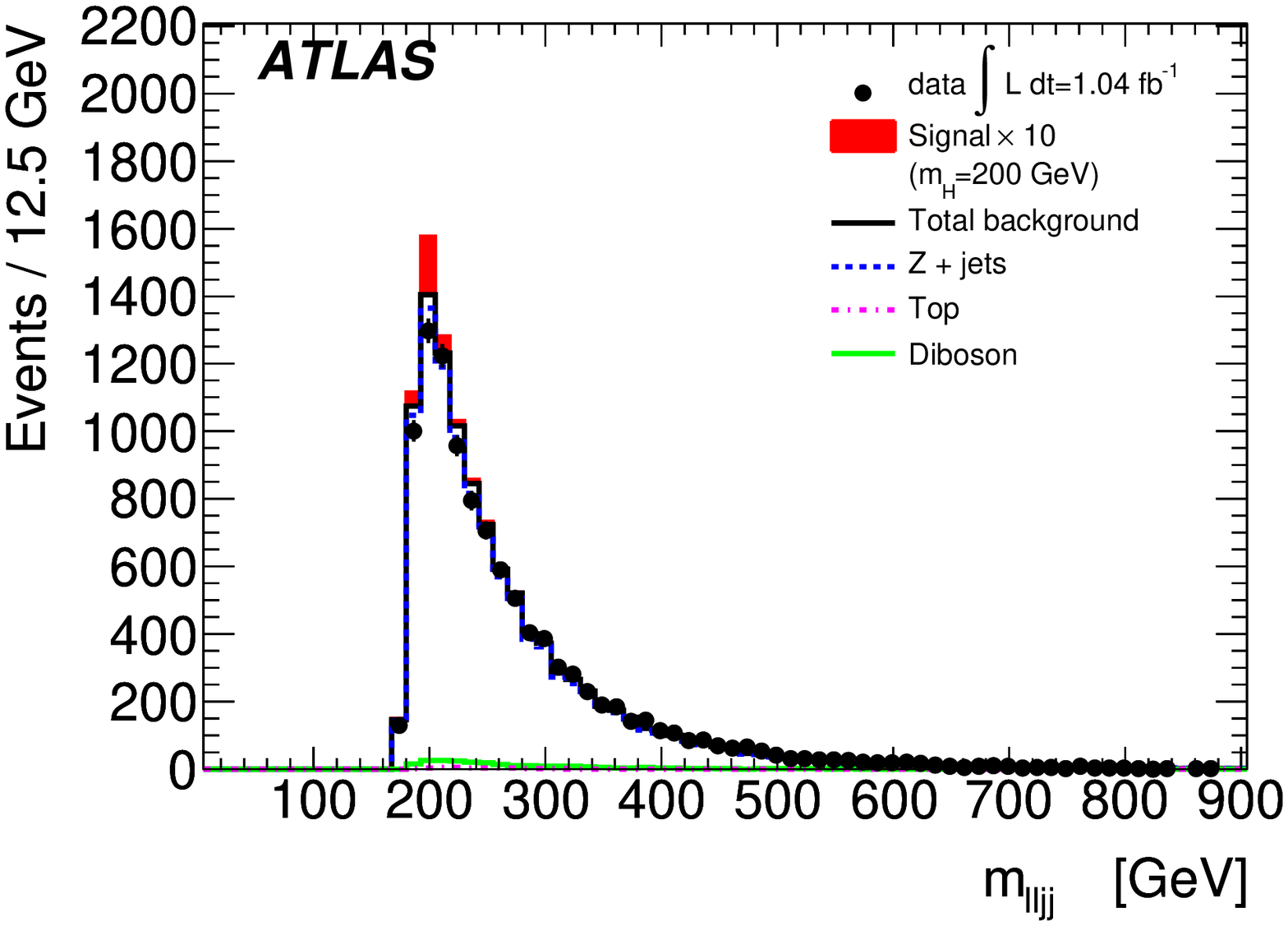}}
    \subfigure[Low-$\mH$, tagged selection.]
{\includegraphics[width=0.45\textwidth]      {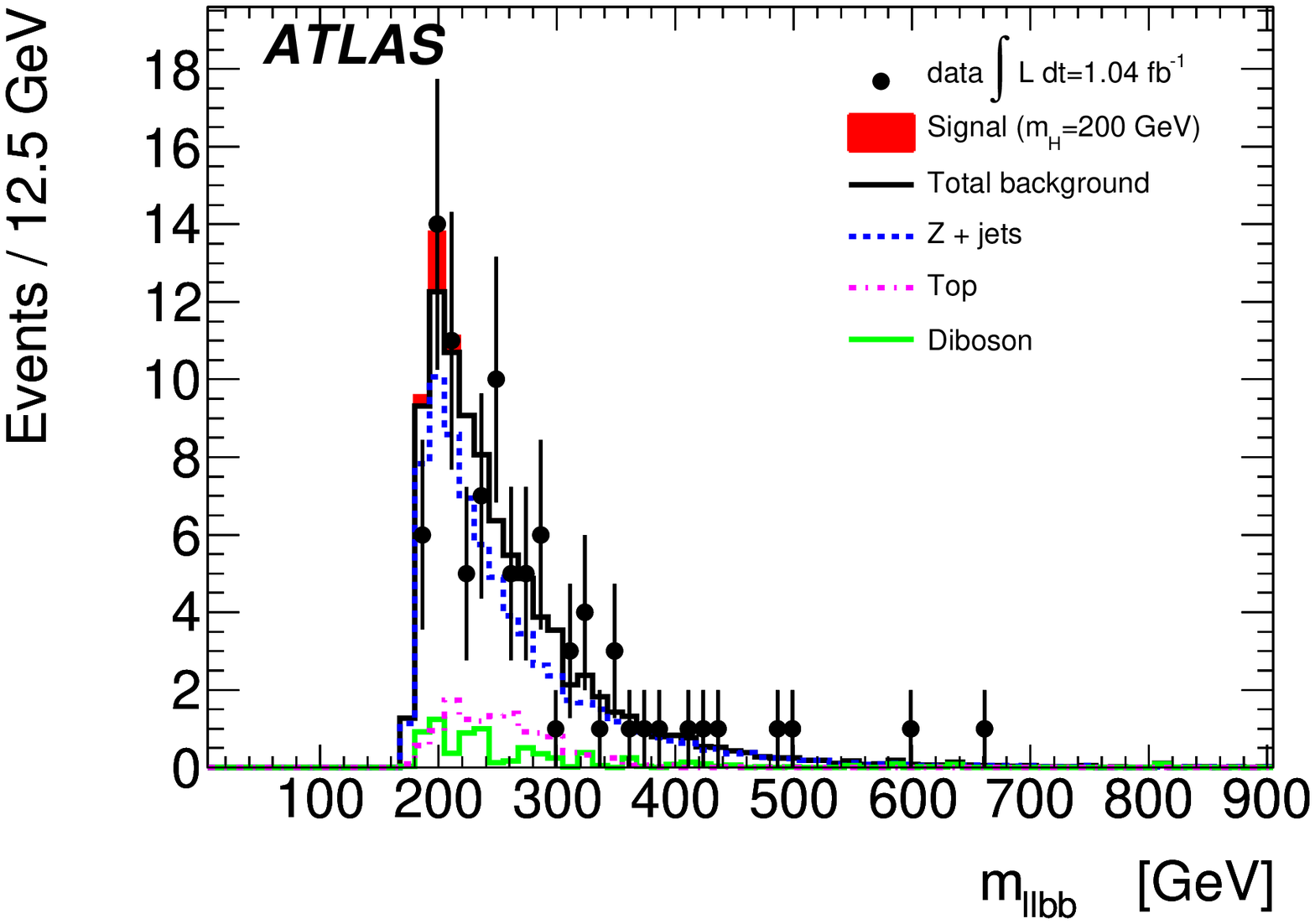}}
    \subfigure[High-$\mH$, untagged selection.]
{\includegraphics[width=0.45\textwidth]      {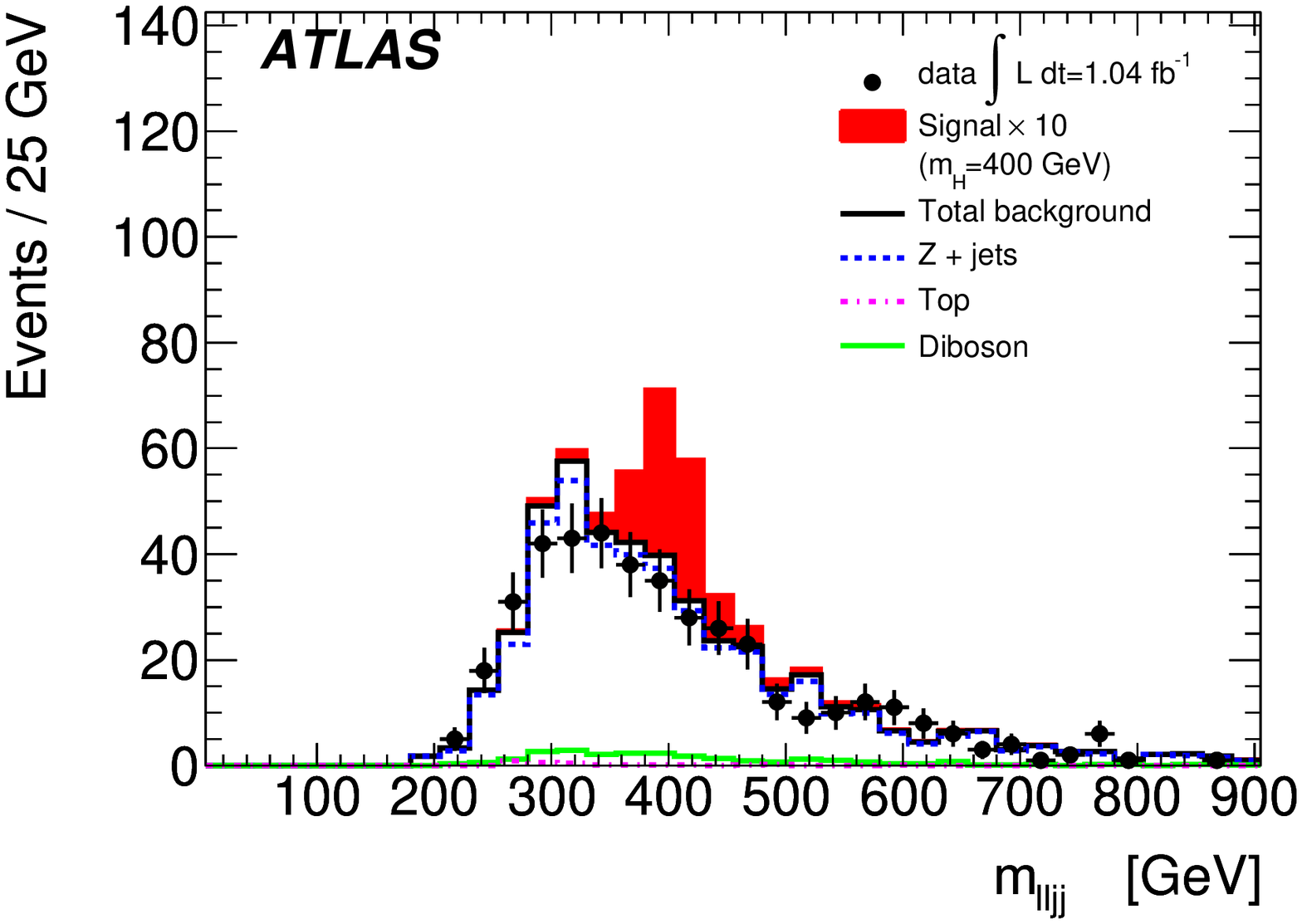}}
    \subfigure[High-$\mH$, tagged selection.]
{\includegraphics[width=0.45\textwidth]      {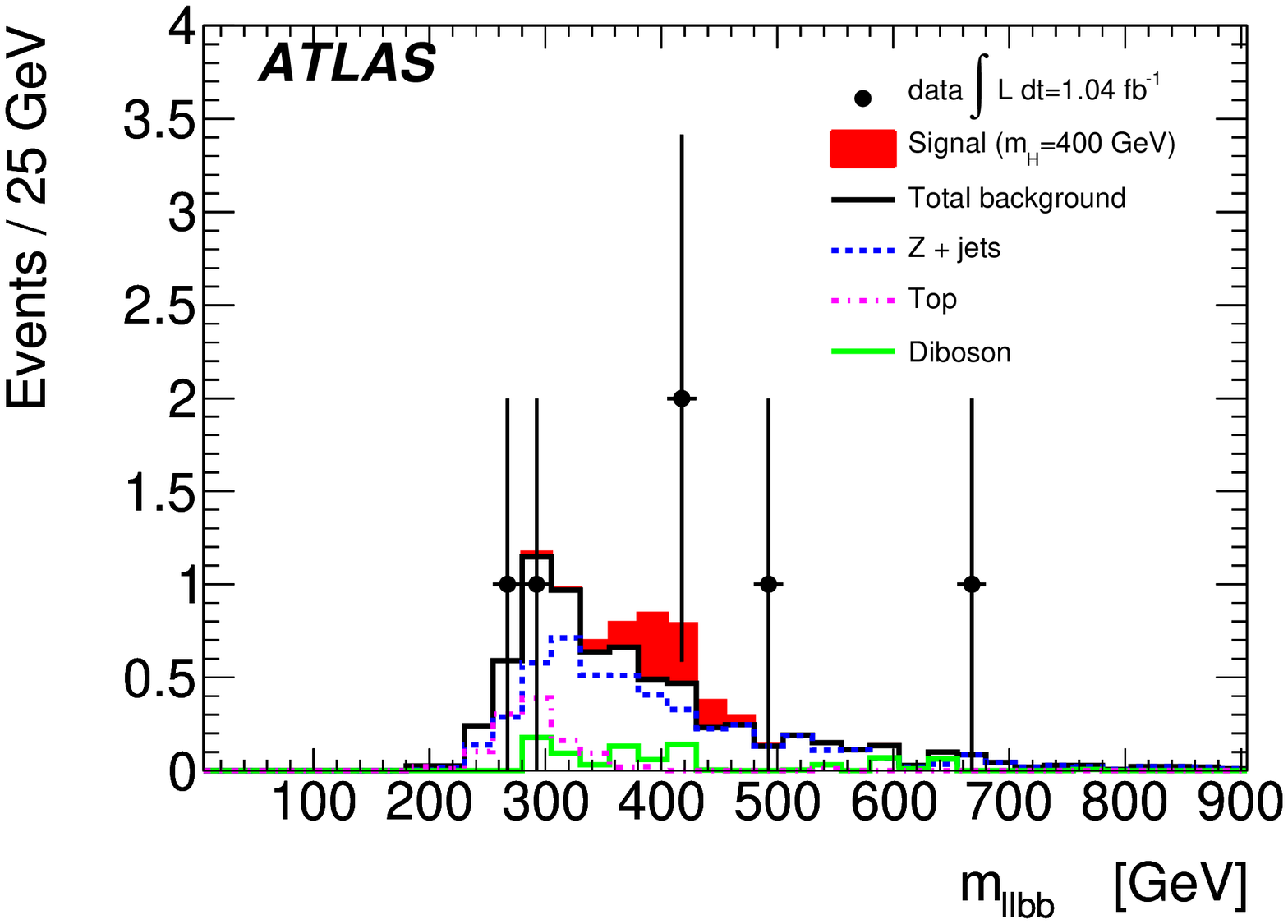}}
  \end{center}
  \caption{The invariant mass of the $\ell\ell jj$ system for both the untagged (a,c) and tagged
    (b,d) channels, for the low-$\mH$ (top row) and high-$\mH$ (bottom row) selections.  Examples
    of the expected Higgs~boson signal for $\mH=200$ and $400\gev$ are also shown; in the untagged
    plots, the signal has been scaled up by a factor of 10 to make it more visible.}%
}
\begin{figure}
  \figresults
  \label{fig:results}
\end{figure}

No significant excess of events above the expected background is observed.
Upper limits are set on the SM Higgs~boson cross section at 95\% CL
as a function of mass, using the $CL_s$ modified frequentist
formalism with the profile likelihood test statistic~\cite{Cowan:2010js,cls}.
This is based on a likelihood that compares,
bin-by-bin using Poisson statistics, the observed
$\mlljj$ distribution to either the expected background or the sum of the
expected background and a mass-dependent hypothesized signal.
Systematic uncertainties,
with their correlations, are incorporated
as nuisance parameters, 
and the tagged and untagged channels are combined by forming
the product of their likelihoods.
\Figref{fig:hlimit} shows the resulting upper limit on the cross
section for Higgs~boson production and decay in the channel
$\htollqq$ relative to the prediction of the Standard Model
as a function of the hypothetical Higgs~boson mass.

\def\fighlimit{%
  \centerline{%
\includegraphics[width=0.7\textwidth]{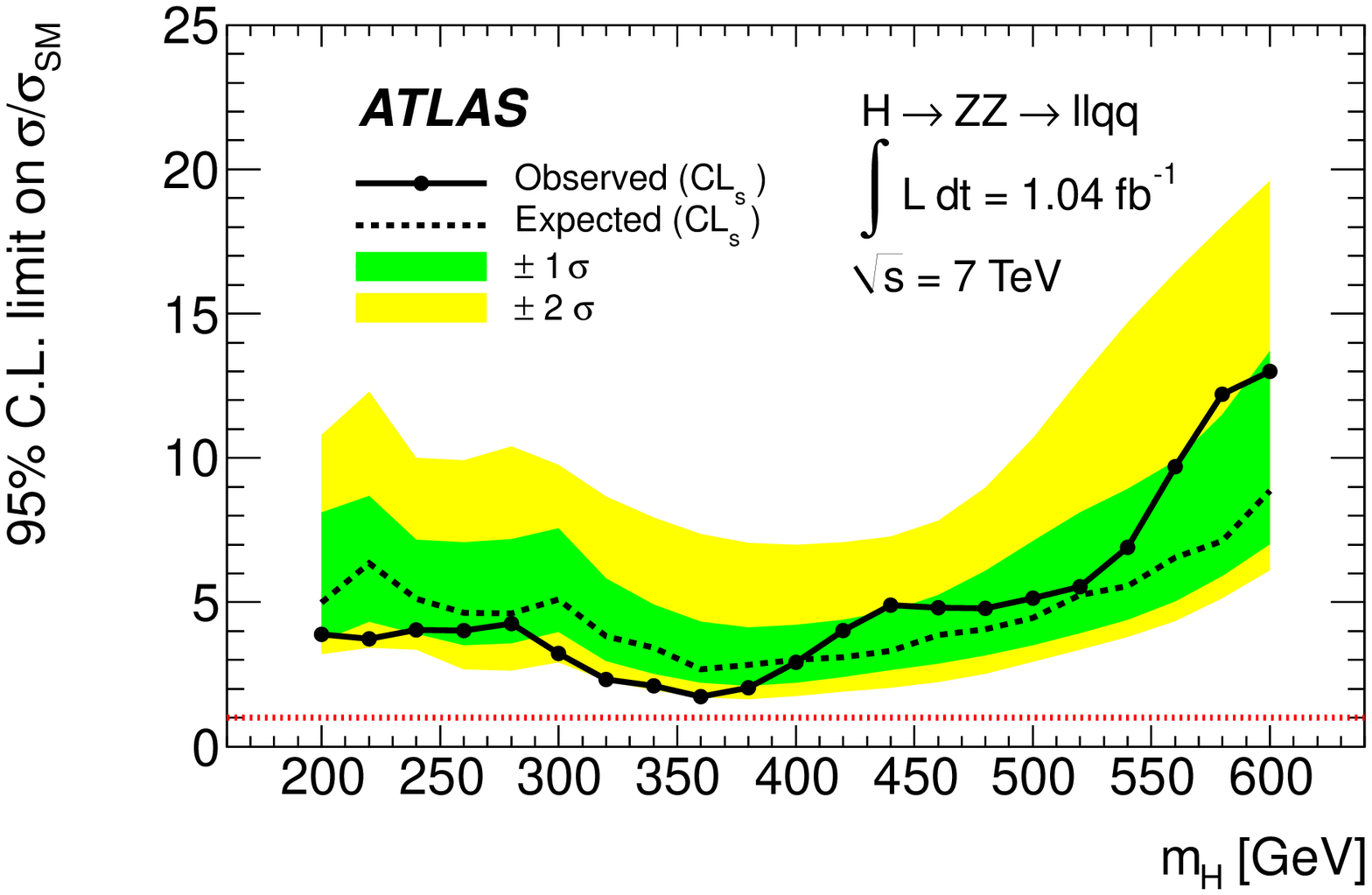}
}%
\caption{The expected (dashed line)
  and observed (solid line) upper limits
  on the total cross section divided by the expected SM
  Higgs~boson cross section, calculated using $CL_s$ at 95\%.
  The green and yellow bands, obtained from interpolating pseudoexperiments,
  indicate the one- and two-sigma ranges in which the limit
  is expected to lie in the absence of a signal.  The dotted red line
  shows the SM value of unity.}%
}
\begin{figure}[htb]
  \fighlimit
\label{fig:hlimit}
\end{figure}

\section{Summary}

A search for the SM Higgs~boson in the decay mode
$\htollqq$ has been performed in the Higgs mass range 200 to $600\gev$
using \lumi\ of $\sqrt{s}=7\tev$~$pp$ data recorded by the ATLAS experiment at
the LHC.  No significant excess over the expected background is found.  With the
present integrated luminosity, there is insufficient sensitivity to
exclude a SM Higgs~boson in this channel at 95\% CL.  The
ratio of the Higgs~boson production
cross section upper limits reported here to the SM
Higgs~boson production cross section ranges from $1.7$ at
$\mH=360\gev$ to about $13$ at $\mH=600\gev$.
These limits are the most stringent to date in this channel.

\section{Acknowledgements}

We thank CERN for the very successful operation of the LHC, as well as the
support staff from our institutions without whom ATLAS could not be
operated efficiently.

We acknowledge the support of ANPCyT, Argentina; YerPhI, Armenia; ARC,
Australia; BMWF, Austria; ANAS, Azerbaijan; SSTC, Belarus; CNPq and FAPESP,
Brazil; NSERC, NRC and CFI, Canada; CERN; CONICYT, Chile; CAS, MOST and
NSFC, China; COLCIENCIAS, Colombia; MSMT CR, MPO CR and VSC CR, Czech
Republic; DNRF, DNSRC and Lundbeck Foundation, Denmark; ARTEMIS, European
Union; IN2P3-CNRS, CEA-DSM/IRFU, France; GNAS, Georgia; BMBF, DFG, HGF, MPG
and AvH Foundation, Germany; GSRT, Greece; ISF, MINERVA, GIF, DIP and
Benoziyo Center, Israel; INFN, Italy; MEXT and JSPS, Japan; CNRST, Morocco;
FOM and NWO, Netherlands; RCN, Norway; MNiSW, Poland; GRICES and FCT,
Portugal; MERYS (MECTS), Romania; MES of Russia and ROSATOM, Russian
Federation; JINR; MSTD, Serbia; MSSR, Slovakia; ARRS and MVZT, Slovenia;
DST/NRF, South Africa; MICINN, Spain; SRC and Wallenberg Foundation,
Sweden; SER, SNSF and Cantons of Bern and Geneva, Switzerland; NSC, Taiwan;
TAEK, Turkey; STFC, the Royal Society and Leverhulme Trust, United Kingdom;
DOE and NSF, United States of America.

The crucial computing support from all WLCG partners is acknowledged
gratefully, in particular from CERN and the ATLAS Tier-1 facilities at
TRIUMF (Canada), NDGF (Denmark, Norway, Sweden), CC-IN2P3 (France),
KIT/GridKA (Germany), INFN-CNAF (Italy), NL-T1 (Netherlands), PIC (Spain),
ASGC (Taiwan), RAL (UK) and BNL (USA) and in the Tier-2 facilities
worldwide.

\bibliographystyle{elsarticle-num}
\bibliography{HllqqPaper}

\clearpage
\input{atlas_authlist}
\end{document}
\bye

%% file: atlas_authlist.tex
\begin{flushleft}
{\Large The ATLAS Collaboration}

\bigskip

G.~Aad$^{\rm 48}$,
B.~Abbott$^{\rm 111}$,
J.~Abdallah$^{\rm 11}$,
A.A.~Abdelalim$^{\rm 49}$,
A.~Abdesselam$^{\rm 118}$,
O.~Abdinov$^{\rm 10}$,
B.~Abi$^{\rm 112}$,
M.~Abolins$^{\rm 88}$,
H.~Abramowicz$^{\rm 153}$,
H.~Abreu$^{\rm 115}$
% ...
\bigskip

% ...
$^{10}$ Institute of Physics, Azerbaijan Academy of Sciences, Baku, Azerbaijan\\
$^{11}$ Institut de F\'isica d'Altes Energies and Departament de F\'isica de la Universitat Aut\`onoma  de Barcelona and ICREA, Barcelona, Spain\\
% ...
$^{48}$ Fakult\"{a}t f\"{u}r Mathematik und Physik, Albert-Ludwigs-Universit\"{a}t, Freiburg i.Br., Germany\\
$^{49}$ Section de Physique, Universit\'e de Gen\`eve, Geneva, Switzerland\\
% ...
$^{88}$ Department of Physics and Astronomy, Michigan State University, East Lansing MI, United States of America\\
% ...
$^{111}$ Homer L. Dodge Department of Physics and Astronomy, University of Oklahoma, Norman OK, United States of America\\
$^{112}$ Department of Physics, Oklahoma State University, Stillwater OK, United States of America\\
% ...
$^{115}$ LAL, Univ. Paris-Sud and CNRS/IN2P3, Orsay, France\\
% ...
$^{118}$ Department of Physics, Oxford University, Oxford, United Kingdom\\
% ...
$^{153}$ Raymond and Beverly Sackler School of Physics and Astronomy, Tel Aviv University, Tel Aviv, Israel\\
% ...
\end{flushleft}
